\documentclass[prb,amsmath,amssymb,twocolumn,aps,superscriptaddress,showpacs,letterpaper]{revtex4-1}
\usepackage{graphicx,color,txfonts,comment}
\usepackage{mathrsfs}
\usepackage{epsfig}

\newcommand{\al}{\alpha}

\newcommand{\g}{\gamma}
\newcommand{\de}{\delta}

\newcommand{\la}{\lambda}

\newcommand{\p}{\pi}

\newcommand{\s}{\sigma}

\newcommand{\w}{\omega}

\newcommand{\De}{\Delta}

\newcommand{\E}{\Xi}

\def\bsig{{\boldsymbol \sigma}}
\def\bvp{{\boldsymbol \varphi}}
\def\q{{\boldsymbol q}}
\def\k{{\boldsymbol k}}
\def\Q{{\boldsymbol Q}}

\def\J{{\boldsymbol J}}
\def\E{{\boldsymbol E}}
\def\v{{\boldsymbol v_\k}}

\newcommand{\pd}{\partial}

\newcommand{\round}[1]{\left( #1 \right)}
\renewcommand{\square}[1]{\left[ #1 \right]}

\newcommand{\beq}{\begin{equation}}
\newcommand{\eeq}{\end{equation}}
\newcommand{\Beq}{\begin{eqnarray}}
\newcommand{\Eeq}{\end{eqnarray}}
\newcommand{\bml}{\begin{multline}}

\newcommand{\bsp}{\begin{split}}
\newcommand{\esp}{\end{split}}

\newcommand{\nn}{\nonumber}

\DeclareMathOperator{\sech}{sech}

\begin{document}

\title{Nonequilibrium probe of paired electron pockets in the underdoped cuprates}

\author{G.~R.~Boyd}
\affiliation{Condensed Matter Theory Center, Department of Physics, The University of Maryland, College Park, MD 20742-4111, USA}
\author{So~Takei}
\affiliation{Condensed Matter Theory Center, Department of Physics, The University of Maryland, College Park, MD 20742-4111, USA}
\author{Victor~Galitski}
\affiliation{Condensed Matter Theory Center, Department of Physics, The University of Maryland, College Park, MD 20742-4111, USA}
\affiliation{Joint Quantum Institute, The University of Maryland, College Park, MD 20742-4111, USA}

\begin{abstract}
We propose an experimental method that can be used generally to test 
whether the cuprate pseudogap involves precursor pairing that acts to gap out the Fermi surface. 
The proposal involves angular-resolved photoemission spectroscopy (ARPES)
performed in the presence of a transport current driven through the sample. 
We illustrate this proposal with a specific model of the pseudogap 
that contains a phase-incoherent paired electron and unpaired hole Fermi surfaces. 
We show that even a weak current tilts the paired band and reveals parts of the previously gapped electron Fermi surface in ARPES if the
binding energy is smaller but close to the pseudogap. Stronger currents can also reveal the Fermi surface
through direct suppression of pairing. 
The proposed experiment is sufficiently general such that it can be used to reveal putative Fermi surfaces 
that have been reconstructed from other types of periodic order and 
are gapped out due to pairing.
The observation of the predicted phenomena should
help resolve the central question about the existence of pairs in the enigmatic pseudogap regime.


\end{abstract}
\date{\today}
\maketitle 

\section{Introduction}
The experimental and theoretical effort to understand the pseudogap phase of the underdoped cuprate superconductors has lasted for
decades~\cite{TimuskStattrev,NormanPepin,normanrev,RevModPhys.78.17,sebastianrev}. 
Most of the current proposed explanations for the phase can be classified into one of two seemingly separate scenarios. One scenario  
interprets the pseudogap as a superconducting precursor state in which electrons pair incoherently above 
$T_c$~\cite{emerykivelson,SFtheory}, and is supported by some transport~\cite{CorsonetalSF}, Nernst~\cite{OngSF1}, and proximity 
effect~\cite{ProximityPG} experiments. The other links the pseudogap to an ordering phenomenon 
that competes with superconductivity. Evidence for this scenario includes 
angle-resolved photoemission (ARPES)~\cite{ARPES_Marshalletal_PRL96etal,ARPES_Normanetal_Nature98etal,HashimotoARPES,ARPESRMP},
electrical transport~\cite{Liberteetal_NatComm11}, quantum oscillations~\cite{QO1etal,QO3etal,PhysRevLett.100.047003,PhysRevLett.100.047004},
x-ray diffraction~\cite{PhysRevB.78.134526,PhysRevB.84.195101}, neutron scattering studies~\cite{NS2etal,NS1etal,PhysRevB.84.195101}, and STM~\cite{Hoffman18012002,hanagurietal2004,Kohsaka09032007etal,CDWSTMetal,kohsakaetal_Nature08etal}.
Recently, observations that support charge density wave (CDW) ordering have been reported in NMR~\cite{highBNMRetal} and
x-ray scattering~\cite{CDW_Scienceetal,CDW_NatPhysetal,CDW_NatPhys2etal} experiments.
%
Although the two scenarios are typically treated as separate interpretations of the pseudogap, 
the phenomena of paired excitations and competing order are not necessarily mutually exclusive. 
In La-based cuprates, for instance, there is evidence for the simultaneous presence of a competing 
order~\cite{lakeetalnature2002,tranquadaPRB2004,khaykovichetalPRB2005,changetalPRB2008,changetalPRL2009}
and pairing~\cite{OngSF1,ProximityPG}. The coexistence of the two phenomena has also served as a basis for 
some theoretical models of the pseudogap~\cite{chubukovdwaverev,GalitskiSachdev}.

Whether precursor pairing exists in conjunction with competing order in the pseudogap regime is an
important question. Here, we propose and model an experiment that can help resolve this
question by directly testing whether the pseudogap contains any putative (or ``ghost") Fermi surfaces that 
are gapped out {\em specifically} due to pairing. 
The recent quantum oscillation and photoemission experiments on the cuprates give an impetus to investigate 
a pseudogap scenario in which both pairing and competing order are incorporated simultaneously.
While quantum oscillations in the pseudogap regime provide evidence for coherent electron Fermi pockets at finite magnetic fields~\cite{QO2etal},
evidence for such pockets is not observed in photoemission at zero magnetic field~\cite{Nodal3etal,hossainetal2008,TranquadaPRB2010}.
However, 
these observations can be reconciled with a scenario in which parts of the {\em putative} Fermi surfaces evade photoemission detection 
due to a strong pairing gap, which in turn is suppressed by a finite magnetic field. This restores the previously hidden Fermi surfaces and gives rise
to the observed quantum oscillations.  A rigorous microscopic theory for such a scenario was recently developed 
in Ref.~\onlinecite{GalitskiSachdev}, but a direct experimental test of the scenario is still lacking.

In this work, we propose an ARPES experiment performed while a transport
current is driven through the sample. 
We show that an {\em arbitrarily weak} current can shift the quasiparticle spectrum and reveal the hidden paired bands which should appear
as new Fermi surfaces in ARPES. 
The specific way in which the current shifts the spectrum would also be indicative of a gap originating from pairing.
For these weak currents, the heating and the influence of the current-induced magnetic field on the path of the photo-ejected electron should be small. 
Large currents can lead to complete depairing, and this should, in principle, also reveal the hidden Fermi surfaces. 
However, such large currents may be impractical due to Joule heating and magnetic field effects. 
To illustrate our proposal, we apply the method to a particular theoretical model~\cite{GalitskiSachdev}, in which
the pseudogap emerges from a fluctuating critical antiferromagnetic state with paired electron and unpaired hole pockets in the anti-nodal and
nodal regions, respectively. 
We emphasize that the applicability of the proposed method is not limited to the model, 
but can be generally used to test the existence of putative Fermi surfaces that have been reconstructed due to other types of order~\cite{sebastianrev},
including CDW~\cite{CDW_Scienceetal,CDW_NatPhysetal,CDW_NatPhys2etal}, but are
gapped out due to pairing.
The main idea of our proposal may also be useful when considering the application of transport current in conjunction 
with other spectroscopic techniques that probe electronic structure.





\begin{figure}[t]
\centering
\includegraphics*[width=0.7\columnwidth]{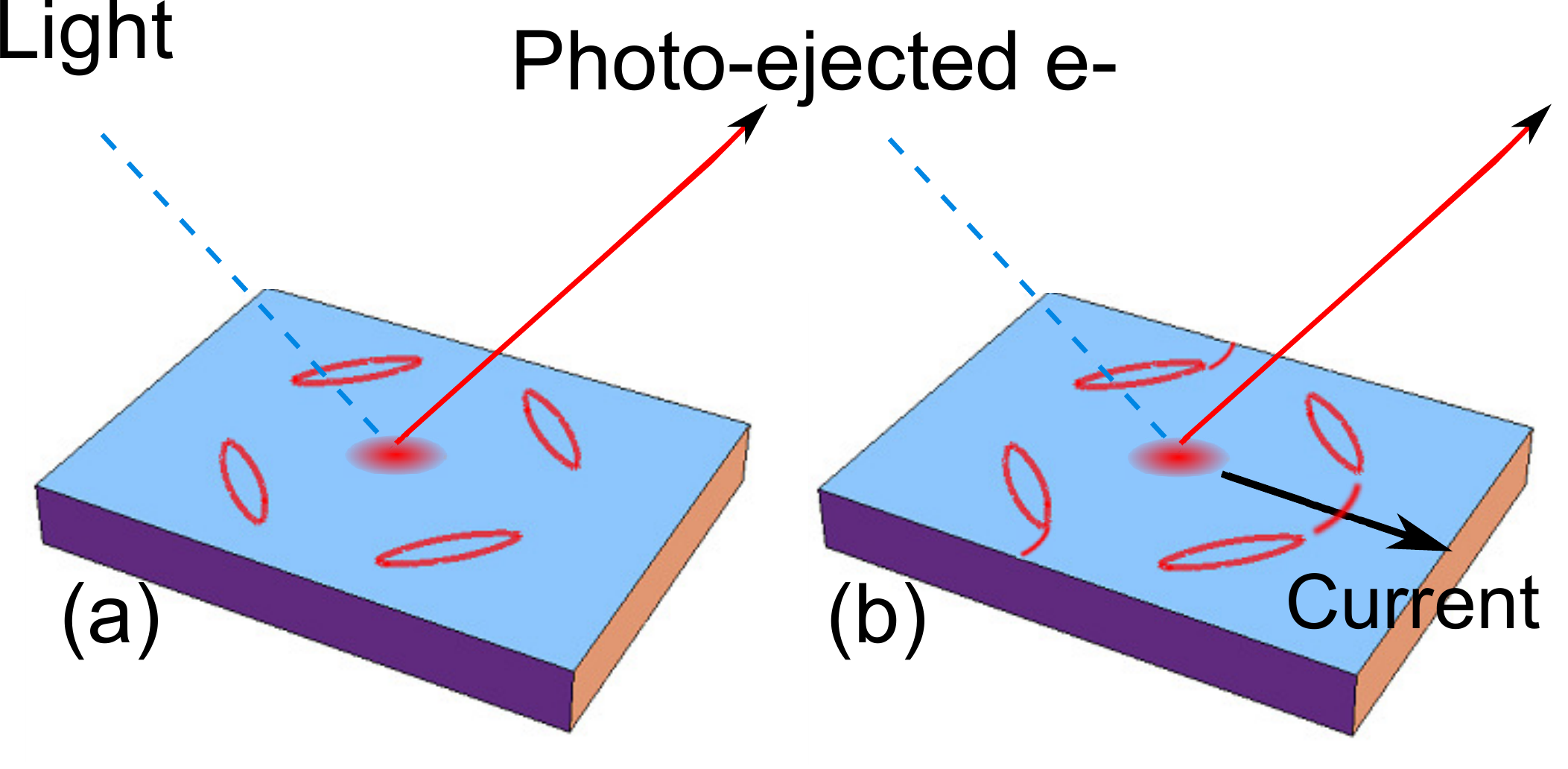}
\caption{(color online) A qualitative illustration of the experimental setup and the predicted phenomenon: (a) the incident light and
ejected electrons that initially leave the electron pockets hidden due to a pairing gap; (b) the electron-pockets 
are partially revealed in the ARPES signal due to a current running through the sample along the anti-nodal direction.
}
\label{ARPEScartoon}
\end{figure}
\section{Model}
To be concrete, we illustrate our proposal using a model of the pseudogap presented in Ref.~\onlinecite{GalitskiSachdev}. 
The model is supported by three recent experimental developments. First, recent work has
observed a proximity-induced pseudogap~\cite{ProximityPG} and supports the idea that the pseudogap is connected to paired quasiparticles.
Second, the experimental discovery~\cite{QO1etal,QO2etal,QO3etal} of small Fermi pockets in the pseudogap phase of underdoped cuprates motivates a
description which incorporates Fermi surface reconstruction. Third, there is a
nodal-anti-nodal dichotomy~\cite{Nodal1etal,Nodal2etal,Nodal3etal,Nodal4etal} observed in
STM~\cite{Yazdanietal} and Raman ~\cite{Ramanetal} experiments, where nodal excitations have an energy
which decreases with decreasing doping, and anti-nodal excitations have a larger energy that increases with decreasing doping.
These key ingredients, namely pairing, Fermi surface reconstruction, and the nodal-anti-nodal dichotomy, are incorporated in the
theory proposed in Ref.~\onlinecite{GalitskiSachdev}. 

According to the theory, the pseudogap phase
emerges from a strongly fluctuating critical antiferromagnetic state with reconstructed Fermi-surfaces consisting of
electron-like pockets in the anti-nodal regions $(\p/a,0)$ and $(0,\p/a)$ and 
hole-like pockets in the nodal regions $(\pm\p/2a,\pm\p/2a)$. 
The components of the {\em physical} electron are described 
in a rotated reference frame set by the local spin-density wave (SDW) order $\bvp=z^*_\s\bsig_{\s\s'}z_{\s'}$,
where the bosonic spinon field $z_\s$ defines the SU(2) rotation.
This parametrization 
gives rise to an emergent gauge field, which 
plays a crucial role in the pairing of the fermions. 
The pseudogap phase is characterized by strongly $s$-wave paired (but uncondensed) electron
pockets and hole pockets~\cite{footnote} that remain unpaired (a weak $p$-wave pairing in the hole pockets is assumed to be suppressed by temperature). 
This is shown to be equivalent to 
$d$-wave pairing when the Brillouin zone is unfolded~\cite{GIL,GalitskiSachdev,RajdeepRVB}. 
Also, the apparent discrepancy between {\em full} pockets in the nodal regions and the Fermi {\em arcs} observed
in ARPES can be reconciled with a model~\cite{Ribhu} that is consistent with what we consider here~\cite{GalitskiSachdev}. 
To reiterate, it is possible that the quantum oscillation experiments~\cite{QO1etal,QO3etal} are observing anti-nodal electron pockets,
which are paired at zero field, but are driven normal by the magnetic field. We show that our proposal can 
falsifiably test this paired electron pocket scenario on Ref.~\onlinecite{GalitskiSachdev} at zero field. 

\section{Theory}
We consider a hole-doped cuprate superconductor in the pseudogap regime subjected to a uniform current $\J$ as shown in
Fig.~\ref{ARPEScartoon}(b). Our main goal is to determine key qualitative features of an ARPES spectrum measured in the presence
of the current. 
A minimal model, which is consistent with the model in Ref.~\onlinecite{GalitskiSachdev}
and  captures the key features necessary to address an ARPES experiment in the presence of current, is
\begin{multline}
\label{H} 
H=\sum_{\k\in RBZ} \sum_{\al=\pm}\square{\xi_{\k f}f^\dag_{\k\al}f_{\k\al}+ \xi_{\k h}h^\dag_{\k\al}h_{\k\al}}\\
-\sum_{\k\in RBZ} \square{\Delta f^\dag_{\k+}f^\dag_{-\k-}+ h.c.}+\frac{\Delta^2}{\la},
\end{multline}
where $f_{\k\pm}$ and $h_{\k\pm}$ are the annihilation operators for the electron-like and hole-like excitations with momentum $\k$, 
respectively, and $\pm$ labels the charge associated with the emergent gauge field. 
We note that $f_{\k\al}$ becomes the physical electron in the SDW ordered state, 
where the spinon field $z_\s$ condenses and the $\pm$ indices become equivalent to the electron spin indices. 
Pairing in the electron pockets is included at the mean-field level by introducing a real $s$-wave pair potential $\Delta$.
The spectra are given by $\xi_{\k f,h}=(\xi_{\k}+\xi_{\k+\Q})/2 \pm[(\xi_{\k}-\xi_{\k+\Q})^2+4\varphi^2]^{1/2}/2$, where
$\xi_\k=-2 t ( \cos k_xa + \cos k_ya ) - 4 t' \cos k_xa \cos k_ya -\mu $. Here, $\Q=(\p/a,\p/a)$, and we will 
take $t'=-0.3t$ and $\mu=-0.6t$~\cite{HashimotoNumbersetal}.
The quantity $\varphi$ is the uniform SDW order parameter, but we stress that $\varphi$ is used here merely to parameterize the underlying
Brillouin zone folding, and that in reality, spin fluctuations are expected to suppress long-range antiferromagnetic order in the pseudogap
phase.
In the above, $\la>0$ is the effective attractive interaction for the electrons generated by the gauge fluctuations. 
Since the pseudogap phase is not characterized by long-range SDW order the attractive interaction mediated by
the gauge fluctuations is, in principle, long-ranged~\cite{GalitskiSachdev}. 

We emphasize here that a calculation
of the current within model (\ref{H}) would give rise to a phase-coherent superflow, which is not correct in the pseudogap regime.
However, the excitation spectrum in the presence of current should
be correctly obtained from these expressions and can reliably calculate the spectral function, which is the central quantity of interest. 

\begin{figure}[t]
\centering
\includegraphics[width=0.47\textwidth]{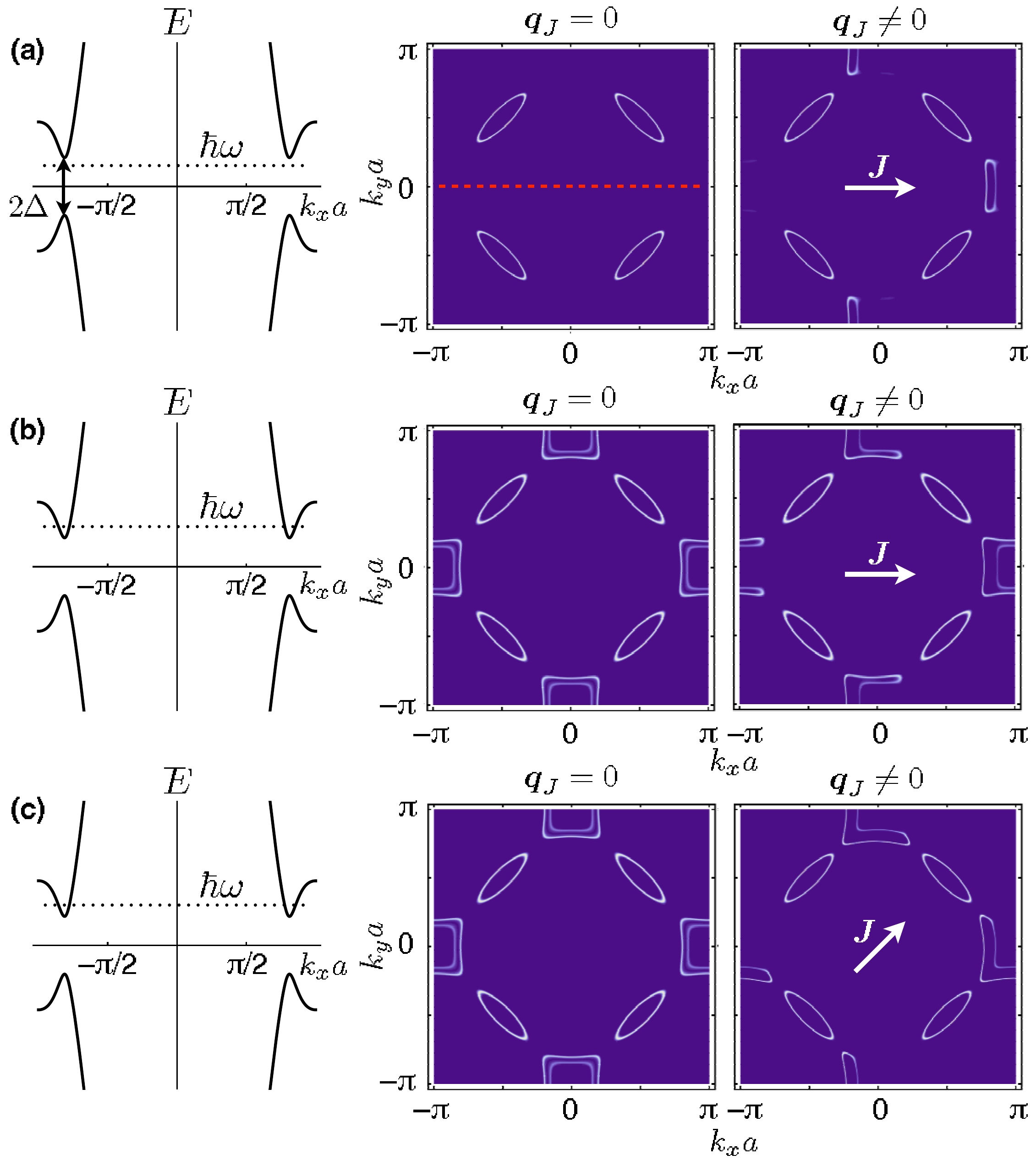}
\caption[]{(color online) Spectral function at zero and finite current for
binding energy $\hbar\w$ (a) just below and (b) above $E_\k$.
As indicated by the white arrows, the current is applied along the anti-nodal direction in (a) and (b), and the nodal direction in (c).
Paired bands on the left are plotted along the red dashed line.
An arbitrarily small $q_J$ can reveal or erase parts of the
electron pockets as long as the binding energy is tuned close to
the edge of a paired band.}
\label{Below}
\end{figure}

\section{Results}
\subsection{Spectral function for a current-carrying pseudogap} 
Transport in the pseudogap can be thought of as a
dissipative flow of holons and charge $2e$ bosons. 
A weak current will only lead to a
simple shift in the holon Fermi surface by an amount $eE\tau_h/\hbar$, where $\tau_h$ is the hole transport relaxation time. 
This does not lead
to a dramatic change in the hole spectral function. As we show below, the presence of even a weak current leads to a noticeable modification in
the spectral function for the electron sector, namely, 
the current can cause sections of the hidden paired electron band to appear or disappear.

The current endows the Cooper pairs with a center-of-mass momentum $\q_\J$, and the single-particle Green function in the presence of a uniform
current can be written as~\cite{makibook,MakiTsuneto}
\begin{equation}
\label{driftG}
\check{G}_\k(i\omega_n)=\left[\hbar(i\omega_n-\v\cdot\q_\J)\check{1}- \xi_{\k f}\check{\tau}_3+\Delta \check{\tau}_1 \right]^{-1},
\end{equation}
where $\w_n=(2n+1)\p k_BT$ are the fermionic Matsubara frequencies, $\v=\pd\xi_{\k f}/\hbar\pd\k$, checks indicate matrices in Nambu space, and
$\check\tau_i$ are the Pauli matrices. The gap is then obtained from the following self-consistent condition,
\begin{align}
\label{GapEq}
1 = \frac{\la}{N}\sum_{\k} \frac{1-n_F(E^+_{\k})-n_F(E_{\k}^-)}{2E_{\k}},
\end{align}
where $n_F$ is the Fermi-Dirac distribution function and $E^\pm_{\k}=E_\k\pm\hbar \v\cdot\q_\J=[\xi_{\k f}^2+\Delta^2]^{1/2}\pm\hbar \v\cdot\q_\J$.
The retarded Green function for the $f_{\k+}$ electrons, $G^R_{\k+}(\w)$, then defines the corresponding spectral function
$A_{\k+}(\w)=-\mbox{Im}G^R_{\k+}(\w)/\p$. One then finds,
\begin{multline}
\label{spectralfunc} A_{\k+}(\w)=s^2_\k\de(\hbar\w-\xi_{\k h})\\+r^2_\k\square{u^2_\k\de(\hbar\w-E^+_\k)+v^2_\k\de(\hbar\w+E^-_\k)},
\end{multline}
where $r^2_\k=[1+(\xi_\k-\xi_{\k+\Q})/(\xi_{\k f}-\xi_{\k h})]/2$ and $s^2_\k=[1-(\xi_\k-\xi_{\k+\Q})/(\xi_{\k f}-\xi_{\k h})]/2$, and the
Bogoliubov coherence factors are given by $u^2_\k=(1+\xi_{\k f}^2/E_\k)/2$ and $v^2_\k=(1-\xi_{\k f}^2/E_\k)/2$. 
Although there can be considerable broadening of the spectral peaks in the pseudogap regime, our proposal should not be heavily 
limited by a lack of resolution because one needs to identify merely the presence or absence of the putative Fermi surfaces.
Therefore, (\ref{spectralfunc}) should correctly
convey the main qualitative effects of the applied current on the spectral function.

\subsection{Spectral function for weak currents} 
From (\ref{spectralfunc}) we see that the spectral function is non-zero when either (i) $\hbar \omega-\hbar\v
\cdot \q_\J-E_{\k}=0$ or (ii) $\hbar\omega-\hbar\v \cdot \q_\J+E_{\k}=0$. The
 effect of the Doppler shift is to move the energy scale $\hbar\w$
away or towards the quasiparticle bands $\pm E_\k$ depending on the sign of $\hbar\v \cdot \q_\J$. 
All that is required here is to use a small $\hbar\v \cdot \q_\J$ to just cross into
either of the $\pm E_{\k}$ bands as long as the binding energy is tuned near the edge of a band. In Fig.~\ref{Below}, we plot the spectral
function (\ref{spectralfunc}) at binding energies just below
[Fig.~\ref{Below}(a)] and above
[Fig.~\ref{Below}(b)] the upper paired band for zero and finite current. We consider the case where the current is applied in the anti-nodal direction,
and also in the nodal direction [Fig.~\ref{Below}(c)].
Plotted on the left are a cut of the paired band dispersion along the $k_y=0$ line in the Brillouin zone (indicated by the red dashed line on
the right in (a)), and the binding energies are indicated by the dotted lines.
In (a), the binding energy is set inside the gap at zero current
and hence the electron pockets are initially not observed. However, once the current is applied, the paired spectrum at $\k$ points where $\v$
is parallel to $\q_\J$ is shifted up in energy while the spectrum at $\k$ points where $\v$ is anti-parallel to $\q_\J$ is shifted down in
energy. This leads to the appearance of some sections of the electron pockets.
In Figs.~\ref{Below}(b),(c) the binding energy crosses both the
hole and the upper paired bands. The application of the current here leads to the disappearance of some sections of the electron pockets.

\begin{figure}[t]
\centering
\includegraphics[width=0.95\columnwidth,angle=0]{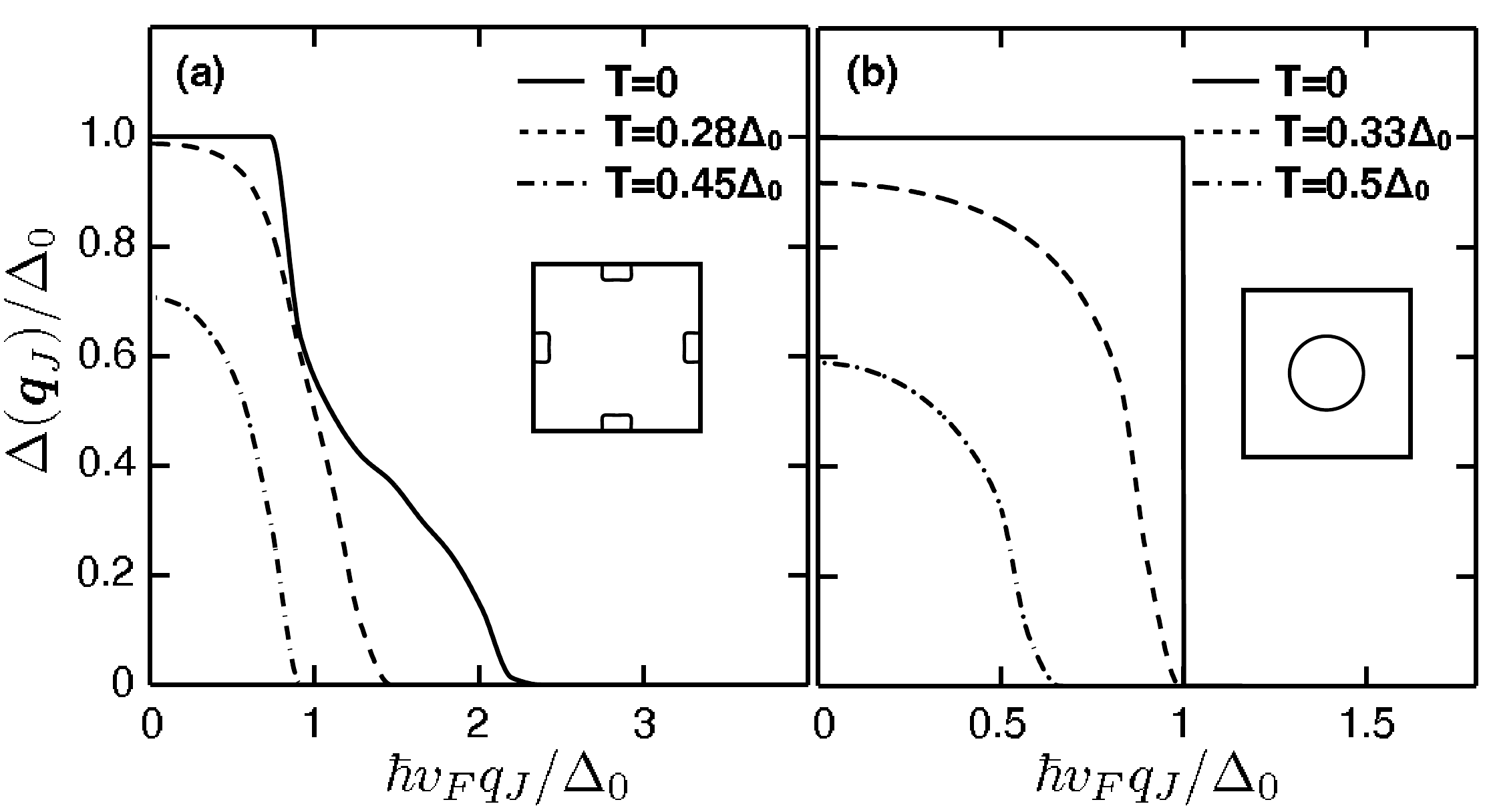}
\caption{Solution to the gap equation (\ref{GapEq}) as a function of $q_J$ along the anti-nodal direction.
In (a), the solution is obtained for electrons with spectrum $\xi_{\k f}$
(with the putative Fermi surface shown in the inset), while in (b) we consider the regular isotropic spectrum $\xi^0_\k=\hbar^2k^2/2m-\mu$ (with the circular putative 
Fermi surface). The gap is normalized with respect to its value at zero temperature and zero current.
}
\label{fig:GapEq}
\end{figure}

\subsection{Depairing due to strong currents} 
A strong current can completely depair the quasiparticles, and this can also reveal the putative electron Fermi surfaces 
in ARPES. The gap solution to (\ref{GapEq}) as a function of $\q_\J$ along the  anti-nodal direction is shown in
Fig.~\ref{fig:GapEq}. As illustrated in the inset, the gap is plotted for the spectrum $\xi_{\k f}$ in (a) and, for comparison, for the regular
parabolic spectrum $\xi^0_\k=\hbar^2k^2/2m-\mu$ in two dimensions in (b). The gap is normalized by $\Delta_0$, which is its value at zero temperature 
and zero current. 
The values $\varphi=0.3t$ and $t/\la=0.7$ were used in (a), and $N_0\la\approx0.34$, where $N_0$ is the density of states at the
Fermi level, was used in (b).

We find that the Fermi velocity for the electron pockets is $\hbar v_F/a\approx 1.3t$. The depairing scale is then be set by
$q^c_J\sim\Delta_0/\hbar v_F$. As we see in Fig.~\ref{fig:GapEq}, at zero temperature, the gap remains robust up to this scale and then shows
a steep or a gradual decrease. 
The shoulder for the $T=0$ result in Fig.~\ref{fig:GapEq}(a) appears because the gap equation (\ref{GapEq}) sums the doppler shift over a 
non-trivial Fermi surface corresponding to the dispersion $\xi_{\k f}$.  
To the best of our knowledge, depairing effects in two-dimensional conventional
superconductors have not been thoroughly investigated, since the initial interest in the early days of BCS theory was in three-dimensional
materials. Many investigations of three dimensional $s$-wave~\cite{BardeenRMP,Nicol1,Nicol}, two dimensional
$d$-wave~\cite{Altman,DegangCurrent} and other exotic pairing~\cite{Khavkine,YongBaekMaki}, however, have been done. 

%

\section{Discussion}
The current-induced magnetic field deflect photo-ejected electrons and can reduce the momentum resolution in the ARPES 
experiment. The deflection can be kept small if the electron collector is well within one cyclotron radius of the sample. 
Assuming an infinite quasi-2D sample with thickness $t$ and a uniform current density $J$, we find the Larmor radius to be
$r=2m_ev_F/(|e|\mu Jt)$, $\mu$ is the permeability. Imposing $d\ll r$, where $d$ is the sample-collector distance, 
this gives an upper bound on $J$, i.e. $J\ll 2mv_F/(|e|\mu dt)$. For $d\sim 1{\rm cm}$, $t\sim1\mu{\rm m}$, and
$v_F\sim250{\rm km/s}$~\cite{NumbersPRBetal}, this gives $J\ll 2\times10^6{\rm A/cm}^2$. 
The upper bound is of order the critical current density
in some cuprate superconductors~\cite{criticalcurr1,criticalcurr2}, 
so this may give a regime where the classical trajectory is minimally deflected but the current
is large enough to shift the dispersion as desired.

Since the pseudogap exhibits dissipative charge current, a potential drop can exist across the incident photon beam spot, and this can compromise
the energy resolution of the ARPES experiment. 
This potential drop can be estimated as $V_{\rm spot}\sim JL/\s_{\rm pg}$, where $J$ is the 2D current density
applied along within the $ab$-plane, $\s_{\rm pg}$ is the $ab$-plane dc conductivity in the pseudogap, and $L$ is the beam spot diameter.
Using $L\sim100\mu{\rm m}$ and $\s_{\rm pg}\sim 5\times 10^5/\Omega {\rm m}$~\cite{Honma2011537}, and requiring that the
energy resolution to be within 1meV, we obtain another upper bound on the 2D current density, i.e.
$J<5\times10^2{\rm A/cm}$. Note that this is an upper bound for a 2D current density along the $ab$-plane.

Heuristically, we now provide the relationship between the drift momentum $\q_\J$ and the current density based on the Drude model. The total
current $\J =\J_b + \J_h$ is composed of holes from the hole pockets $\J_h = \sigma_h \E =(n_h e^2 \tau_h/m_h)\E$ and from uncondensed pairs
$\J_b = \sigma_b \E = (n_b(2e)^2\tau_b/2 m_e)\E$. Here, $n_b$ is the number density of uncondensed pairs and $n_h$ is the density in the hole
pockets. In the pseudogap phase we have a mix of two dissipative fluids whose conductivities will add $\J = (\sigma_b +\sigma_h)\E$. At steady
state, $\q_\J = 2e\E\tau_b $. This means that for a given $\J$ the drift momentum is related to total current via $\q_\J=
2e\J\tau_b/(\sigma_b+\sigma_h)$.

Strictly speaking, the ARPES spectrum must be computed in the physical electron basis obtained  by rotating back
our electron-like excitations using the spinon field $z_\s$. In the absence of long-range N\'eel order, these spinons are uncondensed
and the spinon fluctuations are known to play an important role in the ARPES spectrum~\cite{Ribhu}. 
Nevertheless, the appearance and disappearance of the electron-like Fermi surfaces
found above should have a clear signature in the ARPES spectrum as well. This is because to lowest order the physical electron spectral function
is simply a convolution of the spectral functions of its composite electron-like and spinon particles~\cite{Ribhu}. The appearance or
disappearance of the Fermi surface for the fermion component should then directly affect the convolved spectral function.


\section{Conclusion}
We have shown that the application of a weak current during ARPES in the pseudogap regime provides a falsifiable test of the model proposed in
Ref.~\onlinecite{GalitskiSachdev}, by either revealing sections of the putative electron
pockets that are gapped out due to pairing, or causing them to disappear, depending on the binding energy and current direction. Although a particular model of the pseudogap
based on spin fluctuations was considered here~\cite{GalitskiSachdev}, 
the proposed experiment is also applicable for revealing putative Fermi surfaces that have been reconstructed from other types of order,
such as the charge-density wave~\cite{CDW_Scienceetal,CDW_NatPhysetal,CDW_NatPhys2etal}, but are hidden due to pairing.
We have also briefly considered the possibility to reveal these hidden bands by suppressing the superconductivity
completely by using a strong current.
While the application of current can compromise both momentum and energy resolutions, we have argued that the effects
can be minimized for applied current magnitudes that are not prohibitively small.

\acknowledgments
G.~R.~B. would like to thank B. M. Fregoso for discussions, and S.~T. would like to thank R. Kaul for 
helpful discussions on Ref.~\onlinecite{Ribhu}. 
This research was supported by DOE-BES DESC0001911 (S.~T. and V.~G.).

\end{document}